\documentclass[usegraphicx,usenatbib,useapjfonts,apj]{emulateapj}
\def\gsim{\;\rlap{\lower 2.5pt
 \hbox{$\sim$}}\raise 1.5pt\hbox{$>$}\;}
\def\lsim{\;\rlap{\lower 2.5pt
   \hbox{$\sim$}}\raise 1.5pt\hbox{$<$}\;}
\usepackage{amssymb}

\begin{document}

\newcommand{\vt}{\mbox{\bf {T}}}
\newcommand{\vlss}{\mbox{\bf {T}}_{lss}}
\newcommand{\vtcmb}{\mbox{\bf {T}}_{cmb}}
\newcommand{\vm}{\mbox{\bf {M}}} 
\newcommand{\vn}{\mbox{\bf {N}}} 
\newcommand{\vf}{\mbox{\bf {F}}} 
\newcommand{\1}{\'\i} 
\newcommand{\wmo}{{\bf [Y1]}}
\newcommand{\wmt}{{\bf [Y3]}}

\def\degr{\hbox{$^\circ$}}
\input epsf 
\def\plotone#1{\centering \leavevmode \epsfxsize=\columnwidth \epsfbox{#1}}
\def\plotancho#1{\includegraphics[width=14cm]{#1}}

\title{Oxygen pumping: mapping the reionization epoch with the CMB}

\author{Carlos  Hern\'andez--Monteagudo\altaffilmark{1},
 Zolt\'an Haiman\altaffilmark{2}, Raul Jimenez\altaffilmark{1,3} and Licia Verde\altaffilmark{1,3}}

\altaffiltext{1}{Department of Physics and Astronomy, University of Pennsylvania, Philadelphia, PA 19104, USA; carloshm@astro.upenn.edu, raulj,lverde@physics.upenn.edu}
\altaffiltext{2}{Department of Astronomy, Columbia University, 550 West 120th Street, New York, NY 10027; haiman@astro.columbia.edu}
\altaffiltext{3}{Instituto de F\1sica Te\'orica, UAM/CSIC, Universidad Aut\'onoma de Madrid, Madrid, Spain.}


\begin{abstract}
  We consider the pumping of the $63.2 \mu$m fine structure line of
  neutral OI in the high--redshift intergalactic medium (IGM), in
  analogy with the Wouthuysen--Field effect for the 21cm line of
  cosmic HI.  We show that the soft UV background at $\sim 1300$\AA\
  can affect the population levels, and if a significant fraction of
  the IGM volume is filled with ``fossil HII regions'' containing
  neutral OI, then this can produce a non--negligible spectral
  distortion in the cosmic microwave background (CMB).  OI from
  redshift $z$ is seen in emission at $(1+z)63.2\mu$m, and between
  $7<z<10$ produces a mean spectral distortion of the CMB with a
  $y$--parameter of $y=(10^{-9} - 3\times10^{-8}) (Z/10^{-3}{\rm
  Z_{\odot}}) (I_{UV})$, where $Z$ is the mean metallicity of the IGM
  and $I_{UV}$ is the UV background at $1300$\AA\, in units of
  $10^{-20}$ erg/s/Hz/cm$^2$/sr. Because O is in charge exchange
  equilibrium with H, a measurement of this signature can trace the
  metallicity at the end of the dark ages, prior to the completion of
  cosmic reionization and is complementary to cosmological 21cm
  studies.  While future CMB experiments, such as Planck could
  constrain the metallicity to the $10^{-2} Z_{\odot}$ level, specifically
  designed experiments could potentially achieve a detection. Fluctuations 
  of the distortion on small angular scale may also be detectable.

\end{abstract}

\keywords{ }


\section{Introduction}

The first sources of light that ended the cosmic dark ages are
expected to have reionized the intergalactic medium and polluted it
with metals (see a review by, e.g., \citealt{LoebBarkana}). To use the
metal enrichment as a potential probe of the reionization epoch,
recent studies have concentrated on the neutral oxygen (OI) produced
by the first massive stars \citep{oh2002,bhms}.  The ionization
potential of OI is only 0.2 eV higher than that of hydrogen (HI), and
the two species are in charge exchange equilibrium. Therefore, oxygen
is likely to be highly ionized in regions of the IGM where hydrogen is
ionized.  However, the recombination time for oxygen (as for hydrogen)
is shorter than the Hubble time at $z\gsim 6$, indicating that oxygen
can be neutral even in regions where H has been ionized but where
short--lived ionizing sources have turned off, allowing the region to
recombine \citep{oh2002}. Indeed, \citet{ohhaiman03} show that such
fossil HII regions can occupy $>50\%$ of the volume of the IGM prior
to reionization.  While the metal pollution of these regions is poorly
understood, it is feasible that they contain significant amount of
oxygen with a neutral fraction close to unity.

Previous work has proposed to exploit the scattering of UV photons by
OI, and the corresponding absorption features in the spectra of
quasars -- the OI forest \citep{oh2002}, analogous to the
lower--redshift HI Lyman $\alpha$ forest (e.g.,
\citealt{BeckerOforest06}).  In the case of HI, another interesting
signature from the high--redshift IGM is the 21cm hyperfine structure
line, which is made detectable by UV pumping by the Lyman $\alpha$
background (the so--called Wouthuysen-Field effect; see
\citealt{field58} or \citealt{FurlaOhBriggs} for a recent review).
Here we examine whether a similar effect occurs for OI.

Here we demonstrate that there exist OI lines with the required
features: the fine structure lines of the electronic ground state,
$44.1, 63.2$ and $145.5 \mu$m, can be pumped by $1300$ \AA\, photons
produced by the first stars or black holes, via the Balmer $\alpha$
line of OI.  We compute the spectral distortion in the CMB created by
this effect, and find that it can be as high as $y \sim 10^{-7}$ if
the OI metallicity at $z\sim 7$, just prior to reionization, is $\sim
10^{-2.5}{\rm Z_\odot}$. This distortion could be detectable with
future CMB experiments, and opens the possibility of performing
tomography of the reionization epoch using this effect.  In
combination with 21cm studies, it could yield direct measurements of
the abundance and spatial distribution of metals in the high--redshift
IGM.

\section{Balmer-$\alpha$ pumping of OI fine structure transitions}

The basic criteria for a line $0\leftrightarrow 1$ of a metal species
or their ions to produce an effect analogous to the Wouthuysen--Field
pumping of the HI 21cm line, are as follows: (i) abundant metals in
the IGM in the required ionization state; (ii) the $0\leftrightarrow1$
transition at a frequency suitable for detection, with Einstein
coefficients small enough to allow the line to deviate from
equilibrium with the CMB; (iii) the upper and lower states should be
connected via allowed transitions to another state (hereafter called
state 2), so that they can be ``pumped'' in a two--step process; (iv)
a large enough background flux at the wavelength $\lambda_{02}\approx
\lambda_{12}$ corresponding to the $0\leftrightarrow2$ and
$1\leftrightarrow2$ transitions.  In particular, the last criterion
imposes the constraint $\lambda_{02}\gsim 1215\AA$ on the wavelength
of the UV pumping photons before reionization, since neutral HI
depletes the background at shorter wavelengths \citep{haimanabelrees}.

We selected oxygen for our study because it can be relatively abundant
at high redshifts and because the similarity of its ionization
potential to that of hydrogen (we leave a systematic search through
other metal lines to future work).  The ground state of neutral oxygen is
split into the three fine structure levels of the outermost electrons in the
$n=2$ shell, hereafter denoted as $0$, $1^a$, and $1^b$.  We found that among
the electronic states of oxygen, these are the only one satisfying all
of the criteria above and which photons  would be observed  in the CMB frequency  band.  
The states $0$--$1^a$ and $1^a$--$1^b$ are connected by magnetic
dipolar transitions [OI] $^3P_2 \rightarrow \;^3P_1$ at 63.2 $\mu$m
and $^3P_1 \rightarrow \;^3P_0$ at 145.5 $\mu$m respectively and the
states $0$--$1^b$ are connected by an electric quadrupolar transition
$^3P_2 \rightarrow \;^3P_0$ at 44.1 $\mu$m.  We shall
  consider only lines involving the $0$ state $^3P_2$ as the
  lower level for the transition, since the CMB ambient field is practically unable to
  populate the $^3P_1$ state, and the 145.5 $\mu$m line will be suppressed.  Since these transitions are forbidden, the
spontaneous emission Einstein coefficients ($A_{44.1\mu m} =
1.34\times 10^{-10}$, $A_{63.2\mu m} = 8.91\times 10^{-5}$ s$^{-1}$,
$A_{145.5\mu m}= 1.75\times 10^{-5}$) are much smaller than the
typical values for electric dipole transitions ($\sim 10^8{\rm
  s^{-1}}$), a property shared by the hydrogen 21cm line with
$A_{H21}= 2.85\times 10^{-15}$ s$^{-1}$.

\begin{figure}
\plotone{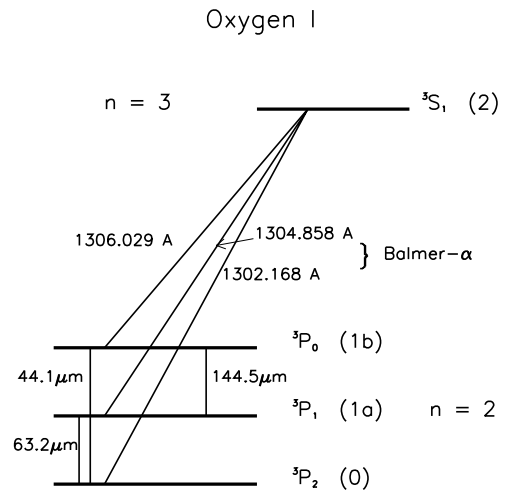}
\caption{Schematic representation of the transitions of neutral oxygen
  considered here. Note the analogy with the 21 cm transition of
  HI. In the 21 cm transition the relevant line corresponds to a
  hyperfine transition, while for OI it corresponds to a fine
  structure transition. The 21cm line is pumped by HI Lyman $\alpha$
  photons, while the OI line is pumped OI Balmer $\alpha$ photons.}
\label{fig:sketch}
\end{figure}

The states $0$ and $1^{a,b}$ are connected to the excited electronic state
$^3S_1$ in the $n=3$ shell (hereafter denoted by $2$) by means of the
absorption of an OI Balmer-$\alpha$ photon with wavelength
$\lambda_{20}\approx\lambda_{21}\approx 1300$\AA. A schematic level
diagram is shown in Figure \ref{fig:sketch}.  In the absence of a UV
field, OI is in thermal equilibrium with the CMB and, as shown in
\citet{bhms}, resonant scattering is more important than
collisionally--induced emission, except at extremely large
over-densities. However, if the first stars or black holes generated a
UV background at 1300\AA, then the relative populations of the fine
structure states 0 and 1 are modified by the two--step ``pumping''
transitions $0\rightarrow 2\rightarrow 1$ and $ 1 \rightarrow
2\rightarrow 0$.

For simplicity we start by considering only one fine structure
transition and only later address the combined effect.  In what
follows we use the indices $i,j$ to denote the two levels in the
fundamental state with $i$ referring to the lowest one. For example
for the 63.2.1 $\mu$m transition ($i,j$) corresponds to ($0,1^a$)
(see fig \ref{fig:sketch}).  The steady state solution for the level
population reads
\begin{equation}
\frac{n_j}{n_i} = \frac{P_{ij}^{UV} + B_{ij}I_{\nu}}
      {P_{ji}^{UV} +  B_{ji}I_{\nu} + A_{ji}}
 \equiv \frac{g_j}{g_i} \exp{\left[-T_{\star,ji}/T_{S,ji}\right]}\,.
\label{eq:sss1}
\end{equation}
This equation defines the spin temperature $T_{S,ji}$. Here, $T_{\star,ji}$
is the equivalent temperature of the $i\leftrightarrow j$ transition
($T_{\star,ji}\equiv h\nu_{ji}/k_B \simeq 231\;(63.2\mu {\rm m}
/ \lambda_{ji})$K with $k_B$ the Boltzmann constant), $A_{ji}$, $B_{ji}$
and $B_{ij}$ are the Einstein coefficients for spontaneous emission,
induced emission and absorption, respectively, $P_{ji}^{UV}$ and
$P_{ij}^{UV}$ denote the UV de-excitation and excitation rates,
$I_{\nu}$ denotes the specific intensity at the resonant frequency of
the $i\leftrightarrow j$ transition, and $g_{1^a}=3$, $g_{1^b}=1$,
$g_0=5$ are the degeneracy factors.  It is useful to define the UV
color temperature, $T_{UV,ji}$ in terms of the ratio of the UV
de-excitation and excitation rates,
\begin{equation}
P_{ij}^{UV}/P_{ji}^{UV}
\equiv (g_j/g_i)\exp[-T_{\star,ji}/T_{UV,ji}]\,,
\end{equation}
which represents the spin temperature reached in the limit of strong
pumping. Following \citet{field58}, these pumping rates can be written
as:
\begin{eqnarray}
P_{ji}^{UV} = & (g_2/g_j)\; \frac{A_{2i}}{\left(\sum_m A_{2m}\right)} A_{2j} 
       \left( \frac{c^2I_{\nu}}{2h\nu^3} \right)_{\nu_{2j}} \\
P_{ij}^{UV} = & (g_2/g_i)\; \frac{A_{2j}}{\left(\sum_m A_{2m}\right)} A_{2i} 
       \left( \frac{c^2I_{\nu}} {2h\nu^3} \right)_{\nu_{2i}}. 
\label{eq:lcoeffs}
\end{eqnarray}
Here $g_2=3$, and the subscript $\nu_{pq}$ in the right parentheses
denotes evaluation at $\nu=\nu_{pq}$, i.e. at the frequency of the
transition connecting states $p$ and $q$.  The sum over $m$ takes into
account the five possible downward transitions from the upper state
$2$, listed in Table (\ref{tab:trans}).

\begin{table}
\begin{centering}
\caption{Possible OI transitions between $n=2$ states and\\the $n=3$ $^3S_1$ level$^{(a)}$.  }
\begin{tabular}{ccc}
\hline
\hline
Wavelength &  $A_{2i}$ & $i$  \\
($\AA$) & ($s^{-1}$) & $\left( ^{2S+1}L_J\right) $ \\
\hline
1302.168 & 3.41e+08 & $\;^3P_2$ \\
1304.858 & 2.03e+08 & $\;^3P_1$ \\
1306.029 & 6.76e+07 & $\;^3P_0$ \\
1641.305 & 1.83e+03 & $\;^1D_2$ \\
2324.738 & 4.61e+00 & $\;^1D_0$ \\
\hline\hline
\end{tabular}
\medskip

(a) Taken from NIST URL site: {\tt http://physics.nist.gov/PhysRefData/ASD/lines\_form.html}.\\
\label{tab:trans}
\end{centering}
\end{table}

From the definition of the UV excitation and de-excitation rates, it
is easy to see that their ratio is proportional to the ratio of the
quantity $c^2I_{\nu}/(2h\nu^3)$ evaluated at two slightly different
frequencies: $\nu_{2j}$ and $\nu_{2i}$. It follows then that the color
temperature can be approximated by
\begin{equation}
T_{UV} \simeq 
\frac{T_{\star}}{-\partial\log{n_{\nu}}/\partial 
                         \nu\mid_{\nu_{2j}}}\frac{1}{ \nu_{ji}}
= \frac{T_{2i}}{3+\alpha_s},
\label{eq:tl}
\end{equation} 
where $n_{\nu} \equiv I_{\nu}/\nu^3$ is proportional to the number of
photons with frequency $\nu$, and $\alpha_s \equiv -d\log I_\nu/d\log
\nu$ is the logarithmic slope of the background spectrum near the OI
Balmer $\alpha$ line. We have assumed that $\partial\log{n_{\nu}}/\partial 
\nu\mid_{\nu_{2j}}\approx \partial \log{n_{\nu}}/\partial 
\nu\mid_{\nu_{2i}} $.

If $\partial\log{n_{\nu}}/\partial \nu\mid_{\nu_{2j}}$ is small and
negative then the pumping mechanism will be more effective in the
transition $i\rightarrow 2$ than in $j\rightarrow 2$, and as a result
the level $i$ will be relatively depopulated.  In particular, the
$j\rightarrow i$ photons will be seen in emission if $T_{UV}>T_{\rm
  CMB}$, which will generally be the case since $T_{2i}\sim 10^5$K
$\gg T_{\rm CMB}$.

The amplitude and shape of the UV photon field at 1300\AA\, at the
beginning of reionization is uncertain. Unlike at the HI Ly $\alpha$
frequency of 1215\AA, however, the high--$z$ IGM should be optically
thin at 1300\AA\, (see discussions of the shape of the soft UV
background in \citealt{haimanreesloeb96,haimanabelrees}).  Therefore,
the relevant background at redshift $z$ should reflect the intrinsic
spectrum of the dominant sources over the $\approx 27\%$ redshift
range $(1+z) < (1+z_{\rm source}) < (1+z)\nu_{\rm HI Lyman \beta}/\nu_{\rm OI
  Balmer \alpha}=1.27 (1+z)$\footnote{Lyman $\alpha$ scattering for
  these sources should be negligible \citep{CM04,FP06}, while the
  contribution from yet more distant sources will be blocked, because
  the photons redshifting down to 1300\AA\ by redshift $z$ will have 
  passed through the HI Lyman $\beta$ resonance (or higher lines), and 
  they will have been broken down into lower energy
  photons.}.  
Over the 0.2\% fractional difference between $\nu_{2j}$
and $\nu_{2i}$, we then expect a relative decrement of $n_{\nu}$ of
$\sim 0.2\% (3+\alpha_s)$ throughout $\nu_{ji}$, or a decrement of
order $\sim 1\%$ for sources with $\alpha_s\sim 2$.  For our
calculations, we adopted a fiducial choice of $\sim 1\%$, but in
practice, our results are insensitive to this value (see below).

Given the specific intensity $I_{\nu}$, the steady state solution for
the level population provides the following expression for the spin
temperature $T_{S,ji}$:
\begin{equation}
\frac{T_{\star,ji}}{T_{S,ji}}= 
\log
\left\{ \frac{1 + \frac{ A_{ji}}{ P_{ji}^{UV}}
          \left[ 1 + \left(I_{\nu}c^2 /2h\nu^3\right)_{\nu_{ji}}\right]}
                                                {\exp{(-T_{\star}/T_{UV})} + 
     \frac{A_{ji}}{P_{ji}^{UV}}\left(I_{\nu}c^2 /2h\nu^3\right)_{\nu_{ji}}}  
                                \right\}
\label{eq:tspin}
\end{equation}
If at a given redshift $z_s$ the presence of dust can be neglected,
$I_{\nu}$ at $\nu_{ji}$ is given by the CMB, $I_{\nu} \equiv
B_{\nu_{ji}}(T_{\rm CMB}[z_s])$.  In the limit of $A_{ji} \gg P_{ji}$, in
equation~(\ref{eq:tspin}) we have $T_{S,ji} \rightarrow T_{\rm CMB}$, whereas
if $A_{ji}\ll P_{ji}$ then $T_{S,ji} \rightarrow T_{UV}$.

As long as the slope of the background at 1300\AA\, is such that
$\alpha_S\sim 1$, equation~(\ref{eq:tl}) implies that $T_{UV} >
T_{\star,ji} > T_{\rm CMB}[z_s]$, so that $T_{S,ji}$ must be bracketed
between $T_{\rm CMB}[z_s]$ and $T_{UV}$.  In the limit of
$T_{\star,ji}\ll T_{UV}$, and thus $\exp[-T_{\star,ji}/T_{UV}]\simeq
1$, and $A_{ji} \gg P_{ji}$ (both of these conditions are satisfied
for the 63.2$\mu$m line), the departure of $T_{S,ji}$ from $T_{\rm
  CMB}$ becomes
\begin{equation}
\frac{T_{S,ji} - T_{\rm CMB}}{T_{\rm CMB}} \simeq \frac{T_{\rm CMB}}{T_{\star,ji}} \frac{g_2}{g_j}
\frac{A_{2j}}{A_{ji}} \frac{A_{2i}}{\sum_m A_{2m}}\left(\frac{2h\nu^3}{c^2I_{\nu}}\right)_{ji} \left(\frac{I_{\nu}c^2}{2h\nu^3}\right)_{2j}\,,
\end{equation}
i.e. it depends linearly on the UV flux at 1300 \AA\, and the
dependence on $T_{UV}$ is negligible.
  
\begin{figure*}
\centering
\plotancho{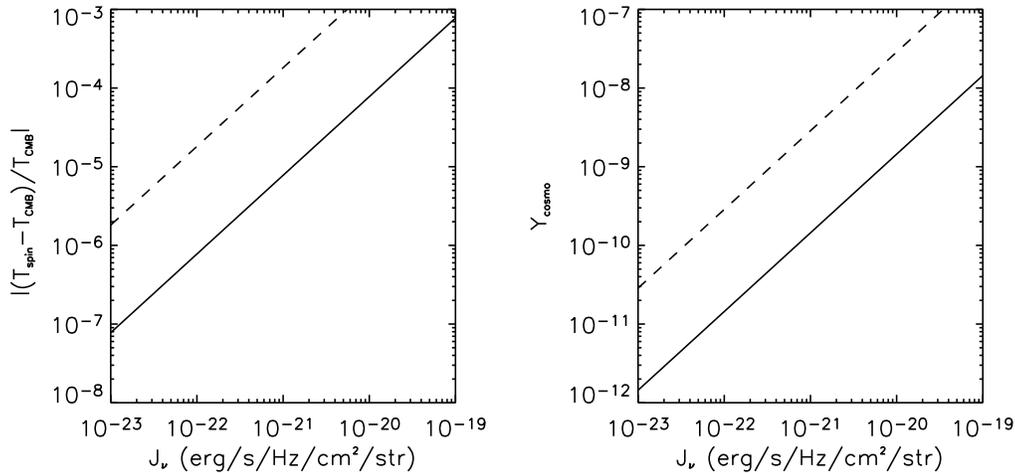}
\caption[fig:ts_y]{{\it (Left)} Relative departure of the spin
  temperature $T_S$ from the CMB temperature as a function of the
  background intensity at the Balmer$\alpha$ frequency of OI.  Note
  that these intensities are within the range of amplitudes required
  for reionization ($\sim 10$ photons per H atom,
  \citet{haimanabelrees}). {\it (Right)} CMB distortion parameter
  $y\equiv \Delta I_{\nu}^{63.2\mu m} / B_{\nu} (T_{\rm CMB} )$ caused
  by the emission of OI in the 63.2 $\mu$m line. In both panels the
  solid line is for $z_s=10$ an the dashed line is for $z_s=7$.  }
\label{fig:ts_y}
\label{fig:fig1}
\end{figure*}

The left panel of Figure~\ref{fig:ts_y} shows $T_S$ as a function of
the UV flux for $z_s=10$ (solid line) and $z_s=7$ (dashed line) 
  for the case of the 63.2 $\mu$m [0 $\rightarrow$ 1a] line.  While
$T_S$ deviates only one part in $10^3-10^7$ from $T_{\rm CMB}$, this
small temperature difference nevertheless modifies the population of
the levels 0 and 1.

----------------------

\vspace{\baselineskip}

\section{Distortions on the CMB due to OI}

The distortion in the CMB spectrum can be parameterized by the
$y$--parameter, $y\equiv \Delta I_{\nu}/B_{\nu}(T_{\rm CMB})$, where
$\Delta I_{\nu}$ is the deviation in surface brightness from the
Planck spectrum with the unmodified CMB temperature.  Before computing
the distortion using the spin temperature above, it is useful to
obtain an order--of--magnitude estimate.  Because the optical depth of
the IGM for the UV photons in our case is small (we find $\tau_{20}=
7.3 \times 10^{-2}$ at redshift $10$ for $Z/{\rm Z_\odot}=10^{-3}$),
multiple scatterings will be rare.  (Note that this is different from
the case of the HI 21cm, where the Lyman$\alpha$ photons responsible
for the pumping have a large optical depth and scatter multiple
times.)  In this case, roughly a fraction $\tau_{20}$ of every
1300\AA\, photon produced by stars will scatter off an OI atom and
produce an ``excess'' $\nu_{ji}\equiv \nu_{j0}$ photon.  Just prior to
reionization, assuming that a few UV photons have been produced per H
atom (corresponding to a background flux of $J_{\nu} = {\rm few}\times
10^{-21}$ erg s$^{-1}$ cm$^{-2}$ sr$^{-1}$ Hz$^{-1}$), the distortion
should be roughly $y \sim f_{\Delta z}^{-1}\tau_{20}\eta\sim
10^{-10}$, where $\eta\sim 6\times 10^{-10}$ is the baryon--to--photon
ratio, and $f_{\Delta z}$ represents the fraction of CMB photon in the
frequency band corresponding to the redshift range $\Delta z$.

To compute the distortion of the CMB at the redshifted $\nu_{ji}$
frequency more precisely, we use the radiative transfer equation
\begin{equation}
\frac{dI_{\nu}}{ds} =  \frac{h\nu}{4\pi}\Psi (\nu ) 
     \bigl[ n_j(A_{ji} + B_{ji}I_{\nu}) - n_i B_{ij}I_{\nu}\bigr],
\label{eq:jnu}
\end{equation}
where $ds$ denotes the cosmological line element along the line of
sight. Since the line profile $\Psi (\nu )$ is narrow, we have ignored
the effect of cosmological expansion.  After taking the OI abundance
in the IGM to be $10^{-3}$ times the solar value \citep{ferraraabel},
we find the optical depth for the $0\rightarrow 1^a$ and
$0\rightarrow 1^b$ transitions
\begin{equation}
\tau_{j0}  =  \frac{c^3}{8\pi\nu_{j0}^3} \frac{A_{j0}}{H(z_s)} \frac{g_j}{g_0} 
\frac{1-\exp{(-T_{\star,j0}/T_{S,j0})}}
{\chi(z_s)} n_{OI}(z_s)
\label{eq:tau_j0}
\end{equation}
to be very small, ($\tau_{1^a0} = 2.1\times 10^{-7}, \tau_{1^b0}\sim 3
\times 10^{-13})$.  In this equation we have approximated the line
profile function $\Psi(\nu )$ by a Dirac delta, and $\chi^{-1}$ is the
fraction of oxygen atoms in the $^3P_2$ level ground level, ($\chi=1$ to
a good approximation). Assuming a constant UV background
(justified by the fact that $\tau_{20} \ll 1$), the net change to the
CMB due to the presence of OI can be written as an integral along the
line of sight of the emissivity, resulting in
\[
\Delta I_{\nu}^{j0}(z_s) = \frac{h\nu_{j0}}{4\pi} \frac{c}{H(z_s)}
 \frac{A_{j0}}{\nu_{j0}}n_{OI}(z_s)\frac{g_j}{g_0} \times
\]
\begin{equation}
\frac{\exp{(-T_{\star,j0}/T_{S,j0})}}{\chi(z_s)}
 \biggl[ 1 - \frac{\exp{(T_{\star,j0}/T_{S,j0})}-1}{\left(2h\nu^3/(c^2I_{\nu}\right)_{\nu_{j0}}}\biggr].
\label{eq:deltaInu}
\end{equation}

The corresponding CMB distortion at the observed frequency
$\nu=\nu_{j0}/(1+z_s)$ is given by $y\equiv \Delta I_{\nu}^{j0}
\;/B_{\nu}(T_{\rm CMB}[z_s])$.  In our case, $\tau_{j0}\ll 1$, and the
distortion parameter can be written in the the simple form $y \simeq
\tau_{j0} [(\exp{x}-1)/(\exp{(T_{\star,j0}/T_{S,j0})}-1) - 1]$, with
$x\equiv T_{\star,j0} / T_{\rm CMB}[z_s]$. 

For the $63.2 \mu$m line $T_{\star,63.2\mu{\rm m}}/T_{\rm CMB} \sim 10$ but
$T_{\star,63.2\mu{\rm m}}/T_{\rm CMB}(T_{S,63.2\mu{\rm m}}/T_{\rm
  CMB}-1) \ll 1$, thus we conclude that
\begin{equation}
y\simeq \tau_{63.2\mu{\rm m}}\frac{T_{\star,63.2\mu{\rm m}}}{T_{S,63.2\mu{\rm m}}}\biggl(\frac{T_{S,63.2\mu{\rm m}}}{T_{\rm CMB}}-1\biggr).
\label{eq:ysimple}
\end{equation}
In this case, the distortion parameter $y$ is simply the relative
deviation of $T_{S,63.2\mu{\rm m}}$ with respect to $T_{\rm CMB}$
multiplied by the optical depth times the ratio $T_{\star,63.2\mu{\rm
    m}}/T_{S,63.2\mu{\rm m}} \sim 10$.

The right panel of Figure~\ref{fig:ts_y} shows the amplitude of the
distortion introduced by the OI 63.2 $\mu$m transition at $z_s= 7$ and
$10$ as a function of the UV background intensity, over a range
expected to be relevant for the epoch just prior to reionization.
Note that, apart from the $T_{\star,63.2\mu{\rm m}}/T_{S,63.2\mu{\rm
    m}}$ factor, eq.~(\ref{eq:ysimple}) is identical to that found for
the HI 21cm spin temperature. Since in our case, for a metallicity of
$Z/{\rm Z_\odot}\sim 10^{-3}$ the relative deviations of
$T_{S,63.2\mu{\rm m}}$ with respect to $T_{\rm CMB}$ are typically
$\sim 10^{2-3}$ smaller than in the HI 21cm scenario, and since
$\tau_{\rm 21 cm}\approx 10^{-4}$,
our $y$--distortion is $\sim 10^{2-3} (Z/10^{-3}{\rm Z_\odot})^{-1}$
smaller than in the 21cm case. For the 44.1 $\mu$m ($0\rightarrow
1^b$) transition, the amplitudes of $y$ are a factor $\sim 5$ larger,
but the relevant frequencies are more contaminated by dust and thus 
more difficult to detect, as we comment next.

\section{Discussion and Conclusions}

Our basic result above is that the OI 63.2$\mu$m (and 44.1$\mu$m)
transition could be seen as deviations in the present--day CMB
black-body spectrum of the order of up to $10^{-8}$ at around
$160-500$ GHz (481 GHz corresponds to z=10 and 160 GHz to z=30, for the 63.2$\mu$m).

The FIRAS experiment has already obtained constraints on the deviation
of the CMB spectrum from a black-body.  We find that in the range of frequencies
corresponding to $10<z_s<20$ for the 63.2$\mu$m line, the FIRAS data \citep{fixenfiras96}
constrains $y<10^{-5}$ at 1-$\sigma$ level, ($4 \times 10^{-5}$ at the
3-$\sigma$ level).
Considering the expected background UV flux at these redshifts we
obtain a FIRAS constraints of 5 (40) solar metallicity at
$z_s>10$. Although this constraint is not yet at an interesting level,
given the measurements of metallicities at $3\lsim z\lsim 5$ (e.g.,
\citealt{schaye,aguirre}), it is the first {\em direct} constraint on
the metallicity of the IGM in the universe at $z\approx 10$.  The
upper limit on the metallicity scales linearly with the constraint on
$y$. For example, by comparing different, accurately calibrated Planck
HFI channels constraints $y < 5 \times 10^{-7}$ could be imposed.  An
experiment able to reduce the FIRAS $y$ uncertainty (e.g.,
\citealt{fixsenmather02}) by two to three orders of magnitude could
impose interesting constraints.

The main limitation to the measurement is foreground dust emission
from the galaxy and from IR galaxies, however clean patches of the sky
(to remove the dust emission from the galaxy) and dust observations at
higher frequencies (to remove the IR galaxies) could be used for this
purpose.  Of course, even if the required S/N is achieved in a future
observation, there could be other cosmological effects (such as
decaying particles; see \citealt{fixsenmather02} for a brief
discussion) producing distortion at the feeble $y\sim 10^{-8}$ level.
It remains to be seen, then, whether the OI distortion can be
disentangled from these.

One further consideration is that clustering of the sources will
increase the signal greatly making it more easily detectable.  The signal
should trace the clustering properties of the fossil HII regions
during reionization and should dominate over the primary CMB angular
power spectrum in the damping tail. The clustering properties of the
signal should be qualitatively similar to those of the
Ostriker-Vishniac effect but will a peculiar frequency dependence: the
signal will be non-zero only for frequencies corresponding to the the
$63.2\mu$m transition at redshifts with non-negligible OI abundance
(and UV background). This will be presented elsewhere.

Our technique complements future experiments that will detect the
H21cm hyperfine transition in absorption because it is sensitive to
different systematics and operates at different wavelengths. 
This would be highly valuable, since HI 21cm + OI measurements in
combination can probe the spatial distribution of metallicity directly.

\vspace{0.5\baselineskip}
ZH acknowledges partial support by NASA through grant NNG04GI88G, by
the NSF through grant AST-0307291, and by the Hungarian Ministry of
Education through a Gy\"orgy B\'ek\'esy Fellowship. CHM, LV and RJ acknowledge support by   NASA grant ADP04-0093, and NSF grant PIRE-0507768.
We thank A. Miller, P. Oh, D. Spergel and J. Miralda-Escud\'e for discussions.

\end{document}